# Photonic switching devices based on semiconductor nanostructures

Chao-Yuan Jin[1,*] and Osamu Wada[2,3]

[1]*COBRA Research Institute, Eindhoven University of Technology, P.O. Box 513, NL-5600MB Eindhoven, the Netherlands*
[2]*CREATE, Kobe University, Kobe 657-8501, Japan*
[3]*JSPS Beijing Office, Library of Chinese Academy of Science, Beijing 100190, China*

**Abstract:**

Focusing and guiding light into semiconductor nanostructures can deliver revolutionary concepts for photonic devices, which offer a practical pathway towards next-generation power-efficient optical networks. In this review, we consider the prospects for photonic switches using semiconductor quantum dots (QDs) and photonic cavities which possess unique properties based on their low dimensionality. The optical nonlinearity of such photonic switches is theoretically analysed by introducing the concept of a field enhancement factor. This approach reveals drastic improvement in both power-density and speed, which is able to overcome the limitations that have beset conventional photonic switches for decades. In addition, the overall power consumption is reduced due to the atom-like nature of QDs as well as the nano-scale footprint of photonic cavities. Based on this theoretical perspective, the current state-of-the-art of QD/cavity switches is reviewed in terms of various optical nonlinearity phenomena which have been utilized to demonstrate photonic switching. Emerging techniques, enabled by cavity nonlinear effects such as wavelength tuning, Purcell-factor tuning and plasmonic effects are also discussed.

## 1. Introduction

The global communication revolution that has been enabled through massive data transmission within optical fiber networks has had a tremendous impact on people's lifestyle and modern industry. Over the past two decades, high-speed optical links have been successfully applied to various systems ranging from long-haul transmission lines to short-haul communications within buildings. However the research focus of photonic integrated circuits (ICs) is moving towards the optical manipulation of information at shorter distances, ultimately on circuit boards or IC chips [1, 2]. Current research in optical interconnection aims to provide an ambitious solution based on optical networks to the physical limitations existing in today's electronic systems, which is believed to be a critical issue for realizing future faster information transfer and processing [3, 4].

One of the key devices to establish a high-bit-rate photonic signal processing system is a photonic switch which functions with an ultrashort delay time [5]. The idea of using optical networks to replace present-day electronic switching fabrics was initially driven by the power hungry nature of the optical to electronic (O/E) and electronic to optical (E/O) conversion. Additionally, there is a speed limitation for the electronic interconnection which typically employs millions of closely spaced metal wires in the state-of-the-art computer systems. Despite impressive advancements in the material and device technology, the current record of the on-chip global interconnection lies at 5 Gb/s per channel [6] whilst the speed record of the chip-to-chip interconnection is at 10 Gb/s per channel [7]. A photonic switch that controls optical signals directly by another light beam with potential recovery times in the pico- or femto-second regimes has the capability for terahertz switching speed. The availability of such a component, combined with the use of low-loss optical waveguides and fibers would provide a promising step towards high-speed and low-cost photonic networks at short distances.

---

[*] E-mail: c.jin@tue.nl

Despite the inherent advantages of the photonic approach there is a major problem in sourcing an optimum nonlinear material to act as the switching medium. For most optical materials, the excitation power required to access to the nonlinear operation regime is fairly high, and the device performance is therefore limited, which represents the well-known problem of the "power/speed trade-off" [8, 9]. To develop optical networks at on-chip or chip-to-chip levels, and especially to reduce the heat dissipation in compact systems, the energy requirement for the operation of individual devices is anticipated to be in a range of 10-100 fJ/bit [10]. Most of the previously proposed devices for photonic switching and modulation would fail this target [11, 12]. The feasibility of optical networks based on photonic switches has thus been frequently debated [13].

Semiconductor nanostructures such as epitaxial quantum dots (QDs) offer the possibility to meet these system goals as a natural consequence of their small volume and atom-like density of states which lead to a high differential gain/absorption [14, 15]. QD based nano-structures are anticipated to generate high optical nonlinearity with ultralow energy consumption. In Ref 15, we attempted to overview the current status of QD-based ultrafast devices which may find ways for their application to future optical networks. From the viewpoint of integrating devices to realize a compact, energy-efficient photonic system, major advantages of QD-based photonic devices include:

- *High energy efficiency.*

  QD ridge-waveguide lasers were demonstrated with a threshold current density of 17 A/cm$^2$ [16]. All-optical QD switches were operated at an energy density of 0.1-1.0 fJ/μm$^2$ [17]. Both values are smaller than those for their quantum well (QW) counterparts.

- *High thermal stability.*

  QD lasers, combined with *p*-type modulation doping technique, achieved temperature insensitivities from 0 to 85$^o$C [18-22]. QD semiconductor optical amplifiers (SOAs) were reported to exhibit ultralow noise figure [23] and negligible pattern effect [24], due to the absence of carrier heating effects in spatially separated QDs.

- *Broad bandwidth.*

  QD-based external cavity lasers can be tuned within a wavelength range larger than 70 nm because of the inhomogeneous broadening of the gain spectra [25]. For the same reason, self-assembled QDs are employed as saturable absorbers for mode-locked lasers [26, 27].

- *Compatibility with an Si-platform.*

  Recent research has proven that QD lasers can be monolithically grown on Ge/Si substrates using a special capability of dot layers to filter out anti-phase boundaries and threading dislocations [28-30].

These achievements provide a solid groundwork for the use of QD-based photonic devices in energy-efficient optical networks which require low-energy consumption, low noise and broadband operation, and compatibility with the Si platform to enable integration with standard electronic ICs. On the other hand, QD-based ultrafast photonic devices can offer operation speeds ranging from 10 to100 GHz or even higher [15]. This initial-stage performance is already comparable with QW-based photonic devices that have been widely employed in present-day photonic ICs.

While a variety of QD-based photonic devices including lasers [31, 32], optical amplifiers [33, 34], optical memories [35, 36] and switches [17, 37] have been successfully demonstrated, their dense integration has been hindered by intrinsic material problems. The two most significant of these are: (1) the inhomogeneous broadening of optical spectra which is a natural consequence of self-assembly, (2) the small cross-section and short interaction length of QDs when interacting with light. These two factors lead to weak optical gain/absorption, which further results in the saturation of differential gain when QD devices are heavily injected and hence limits the nonlinear dynamics [38]. Whilst a great deal of efforts have been devoted to the growth side of QD materials to achieve higher uniformity and higher density, an alternative way to enhance the light-QD interaction and the optical nonlinearity has also been investigated by combining QDs with photonic cavities. Figure 1 illustrates a few examples of QD-based nonlinear photonic devices using micro-discs [39], vertical cavities [40], photonic crystal waveguides/cavities [37], and metallic nanowires/nano-antennas [41]. In the later two cases, not only the active region, i.e. the QD layer, is based on nano-materials, but also the cavities or quasi-cavities (such as metallic nanowires or nanoparticles) are nano-structures. It is therefore essential to treat the QD/cavity combination as a whole system to study its optical nonlinearity. This approach provides a unified view of the underlying physics for most switching devices based on semiconductor nanostructures due to the similarity between QDs and other active nano-materials, and actually brings in a unique opportunity to re-examine the fundamental limitations which have confronted conventional photonic switches for a long time.

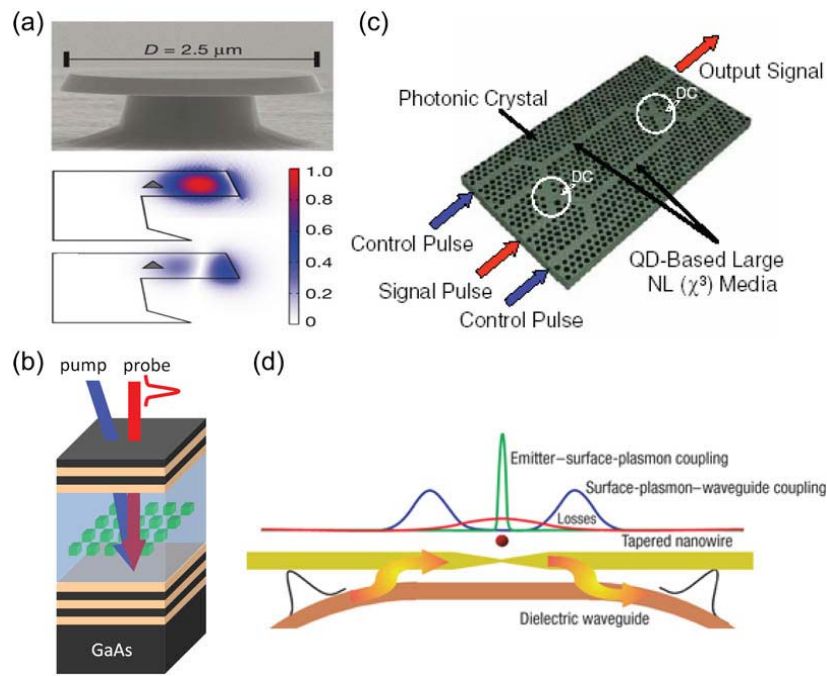

**Figure 1.** QD photonic switches based on four kinds of cavity structures: (a) micro-discs. Reproduced from [39] by permission of *Nature*. (b) vertical cavities. Reproduced from [40] by permission of the American Institute of Physics; (c) photonic crystal waveguides. Reproduced from [37] by the permission of the Institute of Physics; (d) metallic nanowires. Reproduced from [41] by permission of *Nature Physics*.

The contents of this paper are divided as follows: Section 2 discusses physical limitations for the QD/cavity combination, especially for the power/speed related issues. Section 3 describes the current status of the photonic switching devices based on optical nonlinearities originating from the QD/cavity combination. Section 4 discusses the emerging nonlinear functionalities which are enabled by novel photonic cavity structures. Finally, section 5 provides some conclusions and outlook for this work.

## 2. Fundamental limitations

The physical limitations of photonic switches have been a subject debated for many years. The first important discussion of this subject was made by Keyes and Armstrong in the late 1960's [42]. They carried out careful consideration on several known optical effects and concluded that the optical power consumption would be one of the major issues which blocks the utilization of optical logic devices. This argument had remained until the emergence of semiconductor QW technology in the 1980's and the discovery of optical bistability [43-45]. Significantly improved optical nonlinearity was demonstrated by utilizing the quantum confined Stark effect (QCSE) in semiconductors [46], which resulted in a series of successful demonstrations of optical logic devices [47] and large scale photonic integration [48]. In the meantime, the idea of using optical resonators to enhance the switching performance was proposed [49] and experimentally demonstrated [50]. A theoretical examination of the resonator switching was conducted by Fork [51]. He concluded that two technical issues need to be further addressed, namely the small cross-section of the gain media and the low finesse of resonators. Based on further improved device performance of QCSE, the potential architecture for high-performance optical networks was discussed intensively [52, 53]. The physical reasons for advantages of the optical interconnection in comparison to its electrical counterpart became much clearer [54] and power/speed related issues are now the focus of continuing research in nonlinear photonic devices [8-13, 42].

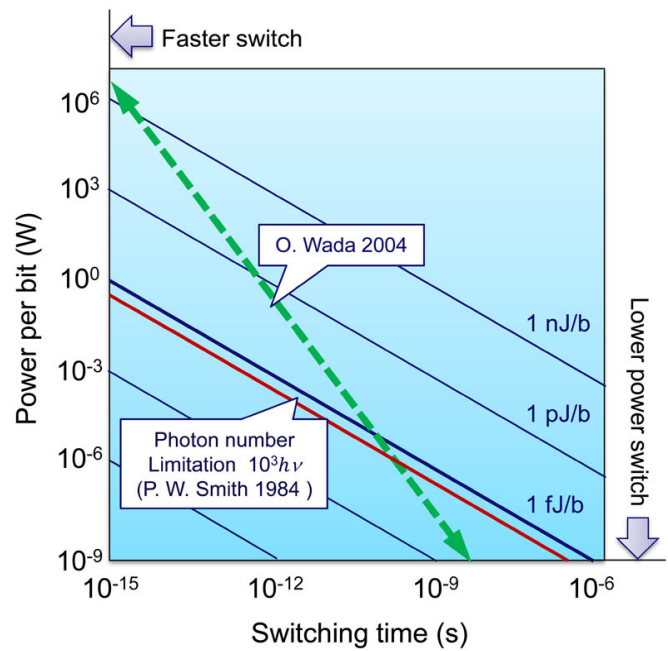

**Figure 2.** A schematic diagram of the "power/speed trade-off" for photonic switches. The red line indicates the photon number limitation based on one thousand photons [8], while the green dashed line is based on a statistic study on the performance of various kinds of photonic switches [5].

Figure 2 summarizes two major limitations of photonic switches, where the data has been extracted from Ref. 5. The first limitation shown by the red curve in the plot was proposed by Smith in 1984 assuming a photon number limitation of one thousand photons [8]. This number was arbitrarily chosen to indicate that a fundamental limit exists in order to achieve a reasonable signal to noise ratio (SNR). This limit can be further connected to the Shannon's Theorem [55]. Very recently, photonic switches based on single QDs have been demonstrated with operation energy down to 8 photons [56]. However, the low-photon-number operation may require a long integration time of optical receivers to enhance the SNR. The exact energy bound related to the photon number limitation will need further discussion. The second limitation depicted by the green dashed line in the figure is based on a summary published by Wada in 2004, which showed the research forefront that had been achieved by various photonic switches based on all-optical, electronic, and optoelectronic means [5]. This straight line indicates the existence of a "power/speed trade-off" for all kinds of switches. Interestingly, the green line has a slope roughly equal to minus two. This slope suggests a limiting relationship between the electronic field (the square root of the switching power) and switching time. A similar formula can be found in Ref. 42 which has been derived as a fundamental bound for optical logic utilizing the nonlinear susceptibility. Although more critical arguments are necessary for the lowest operational photon number, there is an appreciable difference between the two limitations depicted above. This indicates that it should be possible to improve the energy efficiency to the level of fJ or even sub-fJ per bit. This class of devices with ultralow operation power has emerged very recently [17, 57, 58].

As a new generation of semiconductor quantum structures, QDs provide an electronic structure resembling that of real atoms [59]. In the following, we study the QD/cavity combined structure using Maxwell-Bloch equations by treating the single QD as a two-level atomic system [14]. We then attempt to evaluate the performance of nano-photonic switches according to system requirements of the power/speed related issues. It should be noted that, in comparison with the status presented in Ref. 51, although the finesse of today's photonic cavities is significantly advanced by a few orders of magnitudes [60, 61] we still encounter a problem caused by the small cross-section and short interaction length of self-assembled QDs.

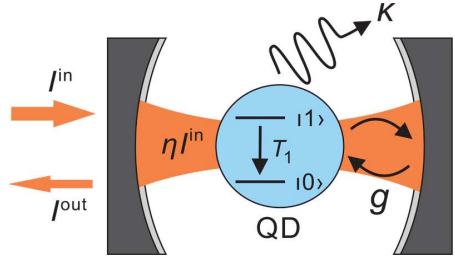

**Figure 3.** A schematic diagram of QDs coupled to an photonic cavity, where $g$ is the coupling strength, $\kappa$ is the cavity loss, $T_1$ is the relaxation time of the dipole, and $\eta$ is the enhancement factor of the input field.

2.1. *Power-density/speed limitation*

The QD-cavity structure can be theoretically represented as coupled oscillators which have been widely employed in cavity quantum electrodynamics (CQED) studies. Figure 3 illustrates a QD positioned in an photonic cavity with a coupling strength $g$, cavity loss rate $\kappa$ and carrier relaxation time $T_1$. The system is assumed to work in the weak coupling regime, i.e. $g \ll \kappa$. The absorption and phase nonlinearity of QDs are derived from the semi-classic approach using Maxwell-Bloch equations [62-64],

$$\frac{d}{dt}\sigma = \left(i\Delta - \frac{1}{T_2}\right)\sigma - ig\delta, \tag{1}$$

$$\frac{d}{dt}\delta = -\frac{\delta - \delta_{eq}}{T_1} + 2i(g\sigma^* - g^*\sigma), \tag{2}$$

where $\sigma$ is the off-diagonal component of the density matrix after neglecting rapidly oscillating terms (it is called the rotating wave approximation), $\delta$ is the population inversion term, $\delta_{eq}$ is the population inversion term at equilibrium, $\Delta$ is the detuning of the optical field from resonance, $g = |\mu||E|/\hbar$ is the QD-field coupling strength, $E$ is the optical field amplitude after neglecting rapidly oscillating terms, and $T_2$ is the dipole dephasing time. The steady-state solution is derived by setting the left-hand sides of the Maxwell-Block equations to be zero. Comparing with the relationship between the polarization and electrical field

$$P = \varepsilon_0 \chi E = N\mu\sigma, \tag{3}$$

the nonlinear susceptibility is obtained

$$\chi = -\frac{c_0 \alpha_0}{\omega} \frac{\Delta T_2 - i}{1 + \Delta^2 T_2^2 + |E|^2/|E_s|^2}, \tag{4}$$

where $\varepsilon_0$ is the permittivity of free space, $N$ is the density of QDs, $\mu$ is the dipole moment, $c_0$ is the speed of light, $\omega$ is the frequency of the dipole moment, $\alpha_0$ is the linear absorption coefficient, and $E_s$ is the saturation electric field. The saturation power density reads

$$I_s = 2\varepsilon_0 c_0 n |E_s|^2 = \frac{\varepsilon_0 c_0 n \hbar^2}{2|\mu|^2 T_1 T_2}, \tag{5}$$

where $n$ is the refractive index. We define a figure of merit by the product of the saturation power density and the characteristic times ($T_1$ and $T_2$) with regard to the power-density/speed limitation

$$\mathbb{F}_{PS} = I_s T_1 T_2 = \frac{\varepsilon_0 c_0 n \hbar^2}{2|\mu|^2}. \tag{6}$$

Here we explicitly use the term "power-density" simply because $I_s$ represents the power density for the saturation electric field. $T_1$ and $T_2$ are the characteristics times which limit the speed of the switching devices. The higher the switching speed becomes, the larger the saturation power density is required. The figure of merit $\mathbb{F}_{PS}$ represents the fundamental physical limit for the fast and power-efficient switching devices. The smaller the figure of merit becomes, the less strict the restriction of the "power-density/speed trade-off" holds. The above equation suggests a constant product between the electric field intensity and the switching time (in case $T_1$ is linearly proportional to $T_2$, i.e. $T_1 \propto T_2$)[§], which matches well with the experimental results as shown in Figure 2. Besides this, a two level system which has a larger dipole moment will have less restriction in the "power-density/speed trade-off". Epitaxial QDs which possess a permanent dipole momentum are therefore good candidates for photonic switching [65]. On the other hand, QDs usually have relatively large values of the dephasing time $T_2$ [66] and the relaxation time $T_1$ [67]. Since $\mathbb{F}_{PS}$ is smaller and the characteristic times are longer compared to those for QWs, photonic switches based on QDs would require lower operation power density.

When a photonic cavity is in resonance with the QD transition, it modifies the spontaneous decay time by the Purcell factor $F$ [68]

$$\frac{T_1^{QD}}{T_1} = F \propto \frac{Q}{V}, \tag{7}$$

---

[§] Only in certain conditions can the relation between the relaxation time $T_1$ and the dephasing time $T_2$ be determined. For example, $T_2$ equals to $2T_1$ for radiatively broadened transitions. Here we use $T_1 \propto T_2$ to simply illustrate that $T_1$ and $T_2$ could have an averaged linear dependence according to the statistic results in Fig. 2.

where $T_1^{\text{QD}}$ is the intrinsic carrier relaxation time of QDs, $Q$ is the cavity quality factor, and $V$ is the cavity mode volume. From equation (6), the cavity switch with Purcell enhancement requires higher saturation power density inside the cavity [69]. However, the input optical field is simultaneously enhanced due to the high-finesse, small-volume cavity, i.e. $I = \eta I^{\text{in}}$ as sketched in Figure 3. The effective figure of merit for the input power and characteristic times becomes

$$\mathbb{F}'_{\text{PS}} = \mathbb{F}_{\text{PS}}/\eta = I_s^{in} T_1 T_2 = \frac{\varepsilon_0 c_0 n \hbar^2}{2\eta |\mu|^2}, \tag{8}$$

where $\mathbb{F}'_{\text{PS}}$ is the new figure of merit for the QD-cavity switch, $\eta$ is the field enhancement factor[§] and $I_s^{in}$ is the input saturation power density. This equation clearly states that the presence of a photonic cavity reduces the power-density/speed limitation by the field enhancement factor of $\eta$. Note that we have not taken into account changes in the group velocity and refractive index in the general description here. For instance, the optical nonlinearity depending on the slow light mode may need a more detailed treatment [70].

In the following, we take two basic examples, the vertical cavity case and the nano-photonic cavity case, as illustrations for the above theoretical description. For a vertical cavity with identical mirrors, ignoring the absorption from QD layers, the field enhancement factor at resonance is equal to [71]

$$\eta_{\text{VC}} = 2\left(\frac{1+R}{1-R}\right), \tag{9}$$

where $R$ is the reflectivity of the distributed Bragg reflector (DBR). The dividing factor of two comes from the standing wave effect. In case of a high-finesse cavity ($R \to 1$), the significant reduction of the power-density/speed limitation is expected. Comparing this to the Q factor of the identical mirror cavity

$$Q = \frac{2\pi L_c}{\lambda} \frac{\sqrt{R}}{1-R}, \tag{10}$$

where $L_c$ is the effective cavity length. Considering that $R \to 1$, the field enhancement factor $\eta_{\text{VC}}$ holds a simple relation using the factor $Q/V$,

$$\eta_{\text{VC}} \propto \frac{Q}{V}, \tag{11}$$

where $V = A_c L_c$ and $A_c$ is the area of the cross section of the cavity.

For the nano-photonic cavity, the situation is much more complicated since the coupling efficiency from guided traveling-wave mode to the cavity mode significantly affects the local field enhancement. We adopt here a simple derivation of the field enhancement factor from the coupled mode theory [72], which gives an intuitive result for both photonic crystal cavities and metallic nano-structures. The electrical field evolvement inside the nano-cavity can be expressed by

$$\frac{dE}{dt} = -\frac{\kappa}{2} E + \sigma E_i, \tag{12}$$

where $\sigma$ is the coupling efficiency from the input wave to the cavity, and $E_i$ is the amplitude of the input optical field. The two optical-field terms in the above equation, $E$ and $E_i$, have excluded the rapidly oscillating terms. From the coupled mode theory [73], $\sigma$ has a simple relation with the radiative part of $\kappa$, $\sigma = \frac{A_c}{\varepsilon_0 V}\sqrt{\frac{\kappa_{rad}}{2}}$ under the definition of $\kappa = \kappa_{rad} + \kappa_{abs}$ where $\kappa_{rad}$ and $\kappa_{abs}$ are the energy decay rates due to radiation and absorption, respectively. The field enhancement factor can be found from the steady-state solution of equation (12)

$$\eta = \frac{|E|^2}{|E_i|^2} = \frac{A_c \kappa_{rad}}{2\varepsilon_0 V(\kappa_{rad}+\kappa_{abs})^2}. \tag{13}$$

---

[§] In this work the field enhancement factor is defined to be the power intensity ratio. There is a difference in the exponent comparing with some of the definitions used in the literature.

For a photonic crystal cavity where $\kappa_{rad} \gg \kappa_{abs}$ and $Q \cong \frac{\pi\lambda}{c_0 \kappa_{rad}}$, the field enhancement factor reads

$$\eta_{\text{PhC}} = \frac{A_c c_0}{2\pi\lambda\varepsilon_0} \cdot \frac{Q}{V} \propto \frac{Q}{V}. \tag{14}$$

For a metallic cavity where $\kappa_{abs} \gg \kappa_{rad}$ and $\cong \frac{\pi\lambda}{c_0 \kappa_{abs}}$,

$$\eta_{\text{MC}} = \frac{c_0^2 A_c \kappa_{rad}}{2\pi^2 \varepsilon_0 \lambda^2} \cdot \frac{Q^2}{V} \propto \frac{Q^2}{V}. \tag{15}$$

It is clear that the figure of merit $\mathbb{F}'_{\text{PS}}$ scales by the Purcell factor $\frac{Q}{V}$ for a photonic crystal cavity, whilst the scaling factor is proportional to $\frac{Q^2}{V}$ for a metallic cavity. Since the two-dimensional (2D) photonic crystal cavity usually has a Q-factor of $10^4$ and a mode volume close to the cubic wavelength, it suggests a possibility of considerable alleviation of the power-density/speed limitation as well as a reduced switching time due to the enhancement of the spontaneous emission rate.

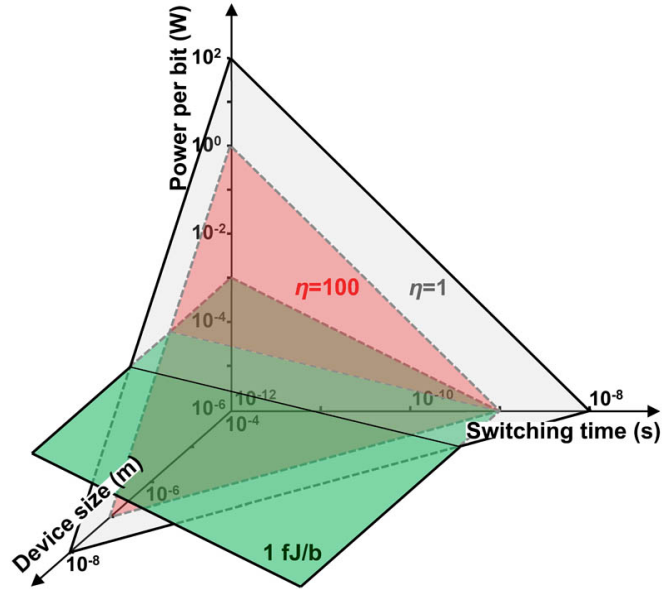

**Figure 4.** A 3D diagram for the power-density/speed limitation. The green plane indicates photonic switches working at 1fJ /bit.

Note that we have used power density in above equations. The total power consumption of the photonic switch highly depends on the device geometry. Photonic switches with nano-sized footprints will consume minimal power and will probably meet the power consumption requirement in the system design. Figure 4 illustrates an updated three-dimensional (3D) version of the 2D "power/speed trade-off" as shown in Figure 2. The parameters used for calculation are the dipole moment for QDs $|\mu| = 7 \times 10^{-29}$ [C·m] and the reflective index $n = 3.3$. We assume a square-shaped cross-section of the photonic switch perpendicular to the light propagation direction where the cross section equals to $d^2$ and $d$ is the device size. Hence the operation power and switching time satisfie $Pt^2/d^2 \approx 10^{-14}$ [kg·s$^{-1}$]. In the figure, the green-colored plane indicates the 1 fJ/bit operation energy contour. The grey and red planes are the constant figure of merit contours indicating "power/speed trade-off" for the field enhancement factors of 1 and 100, respectively. The figure clearly states that when the feature size of photonic switches decreases and the field enhancement factor increases, it is easier to achieve low-energy operation at fJ levels.

*2.2. Enhancement of optical nonlinearity*

It is generally concluded that QD switches require low power to operate since the state filling effect in a single QD uses only one photon to alter the optical spectrum from absorption to transparency. Indeed, an energy consumption less than 1 fJ/bit can be estimated using simple calculations [15]. However, the light passing through QD layers usually experiences low absorption and hence results in low contrast ratio in the switched signal. This effect can be seen clearly by the formula of the third-order nonlinear susceptibility $\chi^{(3)}$. Because $\chi^{(3)}$ is usually proportional to the linear absorption coefficient $\alpha_0$ [62], a weak absorption medium, such as the self-assembled QD ensemble, inevitably corresponds to low optical nonlinearity.

The ratio between the $\chi^{(3)}$ and $\alpha_0$ can be used to define the figure of merit of the third order nonlinearity

$$\mathbb{F}_n = \frac{\chi^{(3)}}{\alpha_0} = \frac{2}{3}\frac{\varepsilon_0 c^2}{\omega}\frac{\Delta T_2 - i}{(1+\Delta^2 T_2^2)^2}\frac{1}{I_s}. \tag{16}$$

Enhanced optical nonlinearity in the cavity structure at large detuning, i.e. $\Delta T_2 \gg 1$, has already been discussed extensively [69, 74]. Here, we only deal with the resonant case, i.e. $\Delta = 0$,

$$\mathbb{F}_n = -i\frac{2}{3}\frac{\varepsilon_0 c^2}{\omega}\frac{1}{I_s}. \tag{17}$$

If we do the same trick as used in the last section, an effective figure of merit of the third-order susceptibility $\chi^{(3)}$ for the input power density of the nano-photonic switch reads,

$$\mathbb{F}'_n = \eta \mathbb{F}_n = -i\frac{2}{3}\frac{\varepsilon_0 c^2}{\omega}\frac{1}{I_s^{in}}. \tag{18}$$

The above equation indicates that the third-order nonlinearity of QDs can be strongly enhanced by the presence of the cavity structure, and the contrast ratio of the switching devices can be improved simultaneously. The enhancement of third-order nonlinearity follows the same expression as shown in equation (11) for the vertical cavity, and in equations (14) and (15) for the nano-photonic cavities, respectively. In other words, the third-order nonlinearity of QDs scales in the same way as the power-density/speed limitation.

For the pump-probe scheme frequently employed in the measurement of photonic switches, while the scaling rule of the power-density/speed limitation suggests the reduced power consumption of the pump beam, the scaling rule of the optical nonlinearity stands for the increased modulation of the probe beam. These two effects can doubly enhance the switching performance in certain configuration as the one we will introduce in the Kerr nonlinearity measurement in Sec.3 [75].

*2.3. Cavity photon lifetime*

Although it has been shown that the cavity enhancement represents positive effects on both aspects of the power-density/speed limitation and the third-order optical nonlinearity, the cavity photon lifetime will be simultaneously increased by the high-finesse cavity and will become an extra factor which limits the cavity enhancement. The cavity photon lifetime is derived as [76]

$$\tau = \frac{1}{2\pi\Delta\nu} \cong \frac{Q}{2\pi\nu} \tag{19}$$

For a cavity with high Q-factor, the cavity photon lifetime should not be larger than the carrier dephasing time of the emitters, i.e. QDs. Otherwise the switching performance will be determined by the photon lifetime in the photonic cavity.

## 3. Current progress on nano-photonic switching devices

The above theoretical analysis indicates photonic switches based on semiconductor nanostructures can potentially meet the energy requirements for future optical networks. The power-density/speed limitation in nano-photonic switches is alleviated by the increase in the Purcell factor and by the decrease of the device footprint. We attempt, in the following, to summarize the recent progress of nano-photonic switch devices exploiting various optical nonlinearities of the QD/cavity combination. Since the vertical cavity represents a simple one-dimensional example of photonic cavities that can be understood analytically, we will derive all formulas necessary for the device design, which evidence the scalability of optical nonlinearity in QD/cavity systems.

*3.1 Absorption nonlinearity*

The absorption nonlinearity of QD materials was initially studied by time-resolved pump-probe spectroscopy [77, 78] and their absorption saturation behavior was confirmed by transient absorption measurement [79]. From that early-stage characterization, the absorbance of QDs was deducted to be $3.3 \times 10^4$ per dot layer in the vertical direction [77] and 1.5 cm$^{-1}$ along the ridge waveguide [79]. The absorbance is relatively small, resulting from the small cross-section and short interaction length of self-assembled QDs. The advantageous side of the QD-based optical nonlinearity is that the figure of merit of the third-order nonlinearity has been reported to be one order of magnitude larger than that of its QW counterpart [79]. According to this, a vertical cavity structure using DBR mirrors was proposed to overcome the small absorbance in QDs and to enhance the QD-light interaction [17, 80].

Figure 1(b) illustrates the operation principle of a reflection-type optical switch using QDs in a vertical cavity. The cavity consists of two DBR mirrors with alternating high- and low-refractive index layers. The layer between two DBR mirrors has a thickness equal to an integer multiple of *λ/2n*, which forms the so-called *λ* cavity. An asymmetric cavity design is actually employed rather than the simple symmetric case described by equations (9) and (10). When the signal (probe) light is injected into the cavity, the light reflected by the front mirror can be completely cancelled by the effective reflection from the back mirror at the cavity resonance where the two reflected light beams are out of phase. This mechanism has been employed in vertical cavity optical switches using QW and bulk materials and is so-called zero reflectivity condition [81, 82]. When a $\lambda$ cavity is considered, the reflectivity at the cavity resonance is derived as [17],

$$R_{cm} = R_F \left[ \frac{1-(R_B/R_F)^{1/2}e^{-\Gamma}}{1-(R_B R_F)^{1/2}e^{-\Gamma}} \right], \tag{20}$$

where $R_F$ and $R_B$ are the power reflectivity of the front and back mirrors at the cavity resonance respectively, $\Gamma = 2\int \alpha(l)dl$ is the integrated absorbance, and $\alpha$ is the absorption coefficient inside the cavity. $\Gamma$ is a non-dimensional parameter which comprises absorbance of the active material inside the cavity. $R_{cm}$ reaches zero when the reflectivity of DBR mirrors satisfies

$$R_F = R_B e^{-2\Gamma}. \tag{21}$$

The light beam at the cavity resonance can fully penetrate into the cavity without reflection at the zero reflectivity condition that is described by equation (21). When strong optical pumping occurs at the cavity resonance, the absorption strength of QDs saturates which induces a violation of the zero reflectivity condition and fulfills the all-optical switching process. Equation (21) suggests an asymmetric cavity design due to the presence of the absorbance term. In QW or bulk materials, large absorption exists in the active medium for the purpose of switching. A fairly small reflectivity is thus required for the front mirror, which leads to a low-finesses cavity [81, 82]. However, the $\Gamma$ value is at the order of $10^{-4}$ in QD structures. A high finesse design needs to be addressed to achieve high Q and high optical nonlinearity, and to compensate for the ultralow absorption. The differential reflectivity at the zero reflectivity condition can be derived by assuming the absorption of the QDs to be saturated, i.e., $\Gamma$ becomes 0,

$$\Delta R_{cm} = R_B \left( \frac{1-e^{-\Gamma}}{1-R_B e^{-\Gamma}} \right)^2. \tag{22}$$

When the back mirror reflectivity approaches to unity, i.e. $R_B \to 1$, the differential reflectivity $\Delta R_{cm}$ reaches its maximum value close to unity. For the asymmetric cavity with weak absorption of QDs, the cavity Q factor can be derived as [71]

$$Q = \frac{2\pi L_c}{\lambda} \frac{\sqrt{R_B}e^{-\Gamma/2}}{1-R_B e^{-\Gamma}} \approx \frac{2\pi L_c}{\lambda} \frac{1}{1-R_B e^{-\Gamma}}. \tag{23}$$

Therefore,

$$\eta \propto \frac{Q}{V} = \frac{Q}{A_c L_c} \propto \frac{1}{1-R_B e^{-\Gamma}}. \tag{24}$$

Comparing equation (22) and (24) at conditions of $\Gamma \to 0$ and $R_B \to 1$

$$\Delta R_{cm} \propto \frac{\Gamma^2}{1-R_B e^{-\Gamma}} \propto \left( \Gamma \cdot \frac{Q}{V} \right)^2, \tag{25}$$

we can clearly see the differential reflectivity of the asymmetric cavity is proportional to the square of the QD absorbance, while the effective absorbance of QDs in a vertical cavity is enhanced by the factor of $Q/V$.

Based on the principles discussed above, the QD/cavity switch has been designed as shown in Figure 5(a). 3×3 layers of self-assembled InAs QDs were inserted into the 3$\lambda$/2 cavity at three anti-node positions of the electrical field. The front (back) mirror consists of 16 (30) pair of GaAs/Al$_{0.8}$Ga$_{0.2}$As layers. Twenty percent Ga was added into the AlGaAs layer to prevent lateral oxidization of Al atoms. The thicknesses of the GaAs and AlGaAs layers were chosen to be 89 and 102 nm, respectively, which formed the cavity resonant mode at 1240 nm. The spatial distribution of the electric field inside the cavity was calculated by a transfer matrix method. An enhancement by 30 times was theoretically estimated for the optical field inside the cavity. 2.6 monolayer (ML) InAs were deposited within an 8 nm $In_{0.15}Ga_{0.85}As$ QW to form a dot-in-a-well structure with an excited state transition around 1240 nm which matches the cavity resonance. The switching dynamics at the cavity mode was characterized by degenerated pump-probe spectroscopy. Switching time of 20 ps, saturation power of 2.5 fJ/cm$^2$, and the maximum differential reflectivity ($\Delta R/R$) close to 10% have been demonstrated in this device as shown in Figure 5(b-c). Another device design has provided a switching time up to 23 ps, a saturation power of 1.0 fJ/cm$^2$, and the maximum differential reflectivity of 3% [17, 80]. All the switching devices have achieved a tuning range beyond 30 nm. The ultralow saturation power has been shown by the cavity enhanced nonlinearity based on a Q-factor around one thousand. Further progress on the differential reflectivity can be achieved by either increasing the layer number of the QDs or the reflectivity of the DBR mirrors. Another method, according to the above theory, is to directly etch the vertical cavity into mesa structures, which could significantly increase the Q-factor and hence the field enhancement. This allows the vertical-cavity switch to exhibit higher differential reflectivity and in the meantime to keep the fJ level operation power. Besides the low power consumption, several approaches have been proposed to accelerate the absorption dynamics of QD switches down to a few picoseconds using physical mechanisms such as the carrier tunneling to an additional QW [83], or the introduction of non-radiative recombination using various impurity dopants [84, 85].

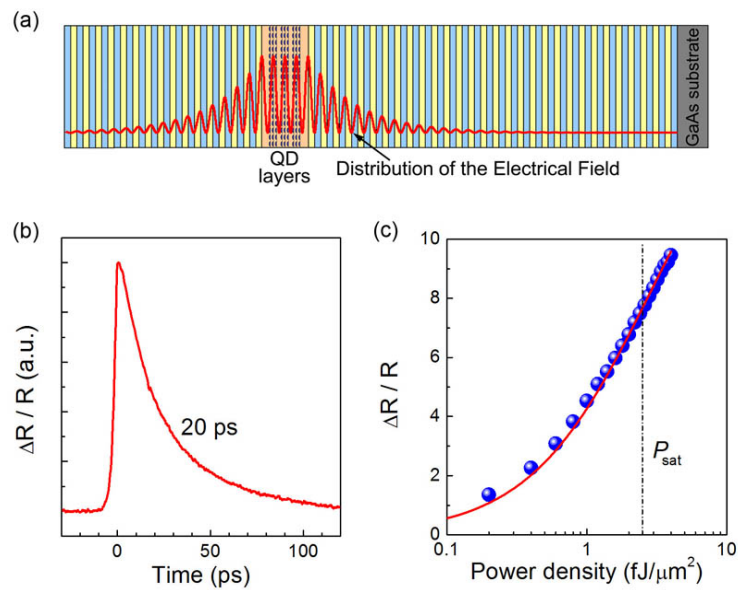

**Figure 5.** (a) A schematic structure of the vertical cavity QD switch. (b) Switching dynamics of an all-optical QD switch. Reproduced from [40] by permission of the American Institute of Physics. (c) The differential refractivity as a function of pumping power density. Reproduced from [40] by permission of the American Institute of Physics.

Very recently, sub-fJ switching has been demonstrated by photonic crystal cavities using InGaAsP/InP QWs [57]. The reported photonic crystal cavity has a Q-factor around 6500 and an ultra-small mode volume of ~0.025 μm$^3$. Extremely high field enhancement can be expected due to the large Purcell factor. For photonic crystal cavities, not only does the large Purcell factor enhancement help to reduce the operation energy, but also the high optical nonlinearity of the refractive index results in nonlinear wavelength shifts. This would start to play a key role in fastening the switching speed and in increasing the differential reflectivity (or transmissivity) when an ultrahigh Q factor is achieved. The nonlinear wavelength shifts in photonic cavities will be discussed in a later section.

When the operation energy reaches the fJ level, the photon number involved in the switching process is only a few thousand, which is close to the limitation proposed by Smith, as shown by the red curve in Fig. 2. However, the forefront of research into low-energy photonic switches has now progressed beyond this limitation. Through efforts to improve the quality of semiconductor-based photonic cavities, the operation energy could eventually approach to the level of a few photons at cryogenic temperature [86]. Initially, strong optical nonlinearity based on single atoms (i.e. QDs) has been proposed for the demonstration of single photon nonlinearity, which relies on the dipole induced transparency and the giant optical nonlinearity in the weak coupling regime ($g \ll \kappa$) [87-89]. The giant optical nonlinearity can be intuitively understood in the framework we described above. When the Purcell factor increases to a certain level, causing an extremely strong field enhancement, the optical reflectivity shows a critical change as the discrete atomic transition gets saturated by a single photon. This opens up the way towards single-photon switches based on semiconductor nanostructures. It can be realized by using single QDs coupled to photonic-crystal cavities [90], vertical pillar cavities [56], and quasi-cavities such as plasmonic nanowires [41]. Interestingly, experimental demonstrations have only succeeded in the strong coupling regime ($g \gg \kappa$), where the physical mechanism differs from the one discussed above. The strong coupling between single QDs and the absorbed single photon inside the photonic cavity splits the atomic transition into dressed states. This blocks subsequent photons to be absorbed at the original energy level. The so-called "photon blockade" represents the underlining mechanism for the optical nonlinearity at the strong coupling regime [91, 92]. Based on the photon blockade, the observation of single photon nonlinearities using semiconductor nanostructures has been reported by many researchers [90, 93-95, 56]. The effective operation energy can reach as low as 8 photons [56]. Note that photonic switching at the level of a few photons has only demonstrated at cryogenic temperature by very long time integration. In that sense, the limitation proposed by Smith has not yet been overcome at practical conditions.

Apart from the absorption nonlinearities, ultrafast gain modulation dynamics using QD-based semiconductor optical amplifiers (QD-SOAs) has been proposed for the application to photonic switching. The optical nonlinearity in QD-based gain media was first studied by time-resolved four wave mixing measurements, and carrier relaxation times at the order of 100 fs have been reported [33, 34]. This was further confirmed by the gain dynamics measurement with sub-ps gain recovery time [96, 97]. Based on the ultrafast optical response, the cross-gain modulation has been achieved with the operation speed up to 160 Gb/s [98]. A 2×2 optical switch array has been fabricated with QD-SOAs, which yields a switching extinction ratio of 24 dB and a transmission rate of 10 Gbit/s [99]. Another example of QD-based switching devices is the semiconductor saturable absorber mirrors (SESAMs). Since QD-SESAMs is not for the purpose of switching networks, we are not going to give a detailed discussion here, an introduction on QD-SESAMs and mode-locked QD lasers can be found in Ref. 32.

*3.2 Phase nonlinearity*

Utilizing the optical phase nonlinearity for QD switches was suggested immediately after the proposal of QD absorption nonlinearity. A π/2 phase shift was observed in a 1 mm QD waveguide with a pump energy intensity of 30 pJ/μm$^2$ [100]. In another prototype device, a phase shift of 4.2 rad/mW for QD-based Mach-Zehnder (MZ) switching was reported based on a 605-μm-long waveguide phase shifter [101]. Apparently, the required operation energy would be too high for practical application of those QD-based phase shifters. Photonic crystal waveguides were consequently employed in the first demonstration of QD switches featured with a MZ structure [37, 102]. A π phase shift was achieved with a net control pulse energy less than 100 fJ/pulse, which is more than three orders of magnitude lower than that for the ridge waveguide devices using bare QDs, since the field enhancement and slow light effect lead to high optical nonlinearity in photonic crystal waveguides [37]. Multiple pulse operation using QD MZ switch has been recently reported with a pulse repetition rate of 40 GHz [103].

Alternatively, phase shifters using vertical cavities have also been studied with a titled pump scheme as shown in Figure 6(a-b) [40]. In the vertical cavity structure, the absorption saturation occurs at the resonant wavelength of the cavity mode. During the switching process, the QD absorption is temporarily changed within the spectral linewidth, which is determined by the homogenous broadening of the absorption spectra of the QD ensemble. The corresponding modulation of the refractive index in dot layers follows the Kramers-Kronig relation. Assuming that no carrier heating effects occur in the passive QD structure, the refractive index change would be close to zero at the cavity resonance. If the probe light is detuned to a different wavelength away from the resonant wavelength but still within the homogenously broadened linewidth of the QD ensemble, the probe light will then experience an appreciable refractive index change in the dot layer. This change of the refractive index induces phase shifts when the light passes though the whole cavity structure. Assuming a transfer matrix for an approximately quarter-wave layer with a small thickness deviation of $\varepsilon\lambda/2\pi$, which has the form of

$$M_{\lambda/4} = \begin{bmatrix} -\varepsilon & i/n_H \\ in_H & -\varepsilon \end{bmatrix}, \tag{26}$$

where $n_H$ is the refractive index of the GaAs layer and $\varepsilon$ stands for the small phase variation in the $\lambda/4$ layer. For the $m\lambda/2$ cavity, the transfer matrix of the cavity layer becomes

$$M_{m\lambda} = \begin{bmatrix} -1 & -2mi\varepsilon/n_H \\ -2mi\varepsilon n_H & 1 \end{bmatrix}. \tag{27}$$

By writing down the transfer matrix for the whole structure with both the front and back DBR mirrors, the total phase shift experienced by the reflected light is derived as

$$\Delta\phi \cong \tan\Delta\phi = \frac{2m\varepsilon n_H \left(\frac{n_H}{n_L}\right)^{2p}}{1+n_H\left(\frac{n_H}{n_L}\right)^{2p-2q}}, \tag{28}$$

where $n_L$ is the refractive index of the AlGaAs layer, and $p$ and $q$ are the periods of GaAs/AlGaAs for the front and back DBR mirrors. In the case of an asymmetric cavity where $q \gg p$ holds, the total phase shift can be simplified to be [17, 40]

$$\Delta\phi \cong 2m\varepsilon n_H \left(\frac{n_H}{n_L}\right)^{2p}, \tag{29}$$

where $n_H(n_H/n_L)^{2p}$ represents the enhancement of the phase nonlinearity due to the vertical cavity. By increasing the number of periods of the front mirror, $p$, the optical phase nonlinearity of the whole structure can be boosted significantly. It is therefore possible to achieve large phase nonlinearity in the vertical direction for QDs. Because equation (28) is obtained under the condition of $q \gg p$, it is natural to assume $R_B \to 1$.

$$\Delta\phi \cong 2m\varepsilon \frac{1+\sqrt{R_B}e^{-\Gamma}}{1-\sqrt{R_B}e^{-\Gamma}} \cong \frac{4m\varepsilon}{1-R_B e^{-\Gamma}} \propto m\varepsilon \frac{Q}{V}. \tag{30}$$

The phase nonlinearity again follows the scaling rule and is proportional to the Purcell factor.

As sketched in Fig. 6(a-b), a MZ interferometer is fabricated for the evaluation of nonlinear phase shifts. The pump beam is designed to be 20° angled from the direction of the probe beam, leading to 5 nm wavelength detuning between the cavity resonant wavelengths for the pump and probe beams. Because the 130 fs optical pulse provided by the ultrafast laser has a spectral broadening of 20 nm, it covers automatically the 5 nm detuning between the pump and probe light, for which the phase shift can be maximized according to the simulation. The fabricated MZ device module is pictured in Fig. 6(b), which has a compact size of $15\times10$ cm$^2$. The QD-cavity structure, which was described in the last section, was inserted into the upper arm of the MZ module. By adjusting the phase compensator involved in the lower arm, the output power intensity of the MZ module exhibits a sinusoid curve corresponding to bright and dark regions in the interference pattern. By pumping the QD switch with a power density of 0.2 fJ/μm$^2$, the interference pattern is shifted by 9.0 μm, which corresponds to an 18$^\circ$ phase shift. By simple calculation, 1.5 rad/fJ can be aimed with a micron-size switching device (Fig. 6(c-d)) [40].

The ultralow power consumption suggests high phase nonlinearities. To further reduce the operation energy, phase shifters comprising single QDs have been demonstrated based on either photonic crystal cavities [104] or vertical pillar cavities [105] at cryogenic temperature. Considering the operation speed, the phase dynamics in passive QD switches and active QD-SOAs exhibits similar recovery time of a few tens of ps [40, 106], as shown in the inset of Fig. 6(c). Although the MZ switching is not affected by the carrier dynamics in the QDs, the repetition rate of the QD-switches will be limited by the relatively slow recovery time.

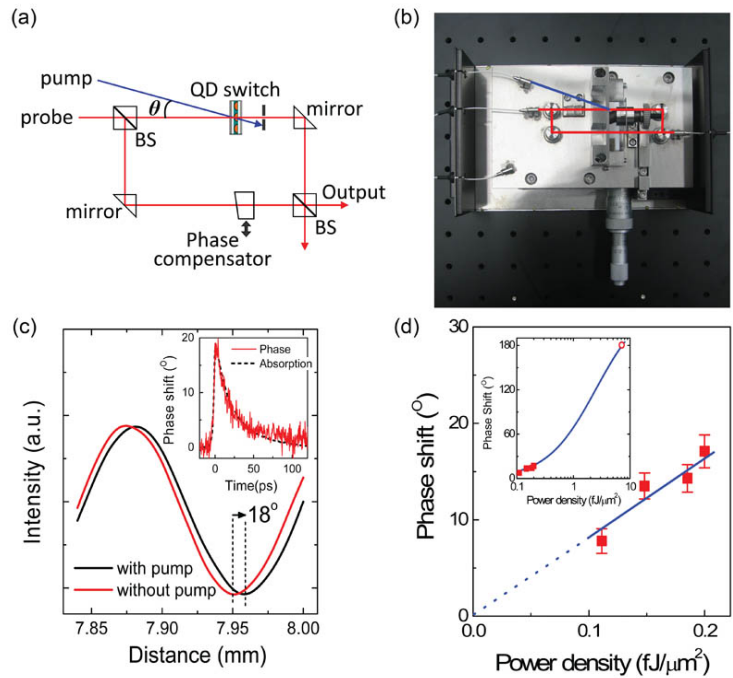

**Figure 6.** (a) A schematic diagram of a Mach-Zehnder interferometer setup for the evaluation of optical phase shifts inside QDs. The pump beam is angled by 20$^\circ$ off the probe beam. BS stands for the beam splitter. (b) A photograph of the Mach-Zehnder module based on vertical cavity QD switches. (c) Interference patterns from the Mach-Zehnder interferometer with and without optical pumping to the QD switching device. The inset shows the phase dynamics (solid curve) compared to the absorption dynamics (dashed curve). (d) Phase shifts as a function of pumping power density. The inset shows simulation results assuming that the refractive index change inside QDs follows the saturation process. Reproduced from [40] by permission of the American Institute of Physics.

*3.3 Kerr nonlinearity*

The Kerr nonlinearity is a quadratic electro-optic effect as the refractive-index change is proportional to the square of the electric field. The enhanced Kerr nonlinearity of photonic crystals was proposed to utilize the off-resonance term in Eq. 16 [69], with the possible trade-off of losing the large field enhancement at resonance. Alternatively, the experimental demonstration of the QD-based Kerr nonlinearity has been achieved with a resonance pumping scheme where a cross-Nicol configuration is applied [107, 108]. Figure 7(a) shows the pump beam is 45° polarized compared to the original polarization of the probe beam along the y-axis. The intensity of the probe beam transmitted through the cavity is recorded, and the x-axis polarization of the transmission is detected as the Kerr signal $I_{kerr}$, which satisfies the following relation:

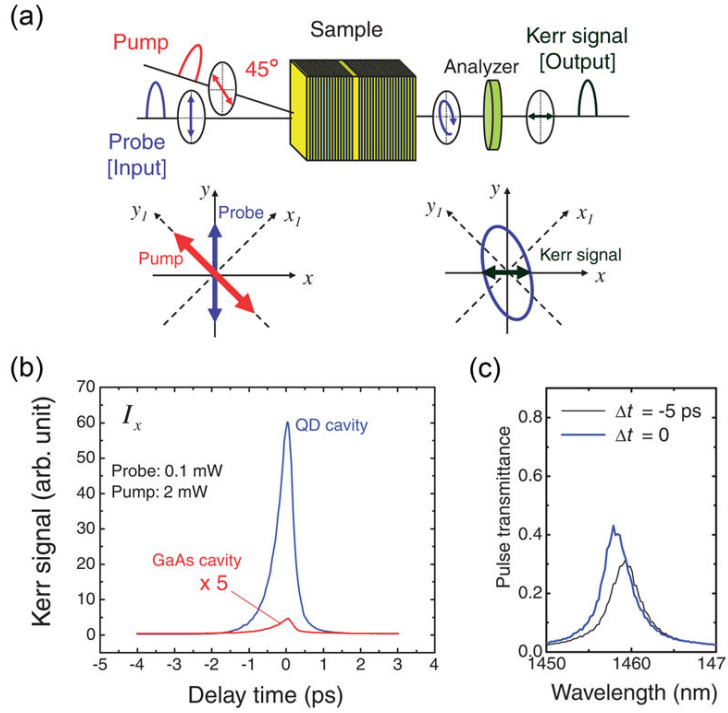

**Figure 7.** (a) A schematic diagram for the Kerr signal measurement of a QD-based vertical cavity switch. The setup follows a cross-Nicol configuration where the pump beam polarization is 45° off the original probe beam polarization. The Kerr signal is recorded from the transmission signal of the probe beam at the polarization perpendicular to the original probe beam polarization. (b) The Kerr signal intensity as a function of the delay time between the pump and probe beams. (c) Transmission spectra of the Kerr signal at the -5 and 0 ps delay. Reproduced from [107] and [108] by permission of The Japan Society of Applied Physics.

$$I_{kerr} \propto \sin^2(\Delta\phi/2) \approx \tfrac{1}{4}\Delta\phi^2, \tag{31}$$

where $\Delta\phi$ is the nonlinear phase shift. By a simple analysis, the nonlinear phase shift reads

$$\Delta\phi = 2\pi n_2 I\, L_c/\lambda, \tag{32}$$

where $n_2 = \frac{3}{2n^2\varepsilon_0 c_0}\chi^{(3)}$ is the nonlinear refractive index, and *I* is the power intensity inside the cavity. Both $\chi^{(3)}$ and *I* have the cavity enhancement by $Q/V$, hence

$$I_{kerr} \propto \left(\frac{Q}{V}\right)^4. \tag{33}$$

For vertical cavities based on the same high- and low-reflective index material combination, the variation in mode-volume among cavities with different layer structures is negligible. It has been thus reported that the Kerr signal is proportional to $Q^4$ in the numerical simulation [75].

The QD cavity switch shown in Fugure 7 has a symmetric cavity design with 13 (13) pairs GaAs/AlAs for the front (back) DBR mirror. The thicknesses of the GaAs and AlAs layers are chosen to be 111 nm and 130 nm, respectively, leading to a cavity resonant mode at 1459 nm. The active region consists of two layers of 3.4 ML InAs QDs embedded in a 167 nm thick $In_{0.35}Ga_{0.65}As$ stain-relaxing layer, which is sandwiched by two $In_{0.35}Al_{0.65}As$ layers. The 35% In allows to tune the QD ground state emission to the 1.55 μm optical communication wavelength. Measurements on the QD switch have shown a significantly improved Kerr signal with an enhancement factor of ~60 in comparison with a control device incorporating bulk GaAs as the active nonliner medium. A reasonable contrast ratio of 10% has been achieved when the switch is on and off. An ultrafast switching time of 0.54 ps is limited only by the photon lifetime inside the cavity, which indicates Kerr switches can potentially operate at THz repetition rates. A parallel work on QW-based vertical-cavity switches using Kerr nonlinearity has achieved a repetition time of 300 fs very recently [109]. However, the exact operation power for the QD Kerr switch has not been reported. Based on the parameters mentioned in Ref 108, there is about a few nJ per pulse penetrating into the vertical cavity. Regardless of the theoretically predicted $Q^4$ enhancement, the operation power for the Kerr switch seems to stay at considerably high level, which supports the existence of intrinsic trade-off between the switching speed and the operation power.

## 4. Cavity nonlinearities for future nano-photonic switching

Although photonic cavities have been intensively studied to enhance the performance of QD switches, the QD/cavity combination remains as a major nonlinear medium for switching devices as we have discussed above. However, the cavity structure will become a dominant source of optical nonlinearity when the device footprint further scales down to nano-scale. Hence we focus in the following three subsections on the nano-cavity side and discuss potentially useful nonlinearities by scanning essential results reported so far. The rich physics at the nano-scale has the prospect not only of enhancing the switching performance but also of enabling new device functionalities in the future.

*4.1 Wavelength tuning of the cavity*

The cavity mode wavelength is usually controlled by design, for example by changing the layer thickness of vertical cavities or by adjusting the lattice constant as well as the filling factor of photonic crystals. The dynamic tuning of the cavity mode provides an additional freedom to enhance the switching contrast at certain wavelengths and to further accelerate the switching speed, which has become a more and more popular tool especially for those photonic switches with low differential reflectivity. Various tuning methods have been demonstrated for photonic cavities, such as thermal tuning [110, 111], free carrier injection [112], liquid infiltration [113], gas condensation [114], mechanical tuning [115, 116], and Kerr effect tuning [117, 118]. Among those tuning mechanisms, the liquid infiltration and gas condensation change the cavity properties permanently or quasi-permanently and hence are not useful for the application to fast photonic switching. The thermal approach achieves a tuning range above 10 nm with a modulation bandwidth less than 1 MHz, which is employed for conventional thermal-optic switches [119]. The mechanical tuning provides large tuning range above 10 nm and with a tuning speed at the MHz level. Recent studies on cavity optomechnics [120] have suggested that GHz coupling rate can be achieved between the optical mode and mechanical mode [121, 122]. High Q cavities combined with optomechanical functions may be potentially useful for certain applications which require modulation bandwidth around GHz. Wavelength tuning based on free carrier injection provides a tuning range of a few nm with a modulation bandwidth of 1~100 GHz and is the most popular tuning mechanism applicable to fast speed photonic switches. The optical Kerr effect represents the fastest tuning speed, only limited by the cavity photon lifetime, but naturally requires high operation power, as discussed in the above section.

*4.2 Purcell factor tuning*

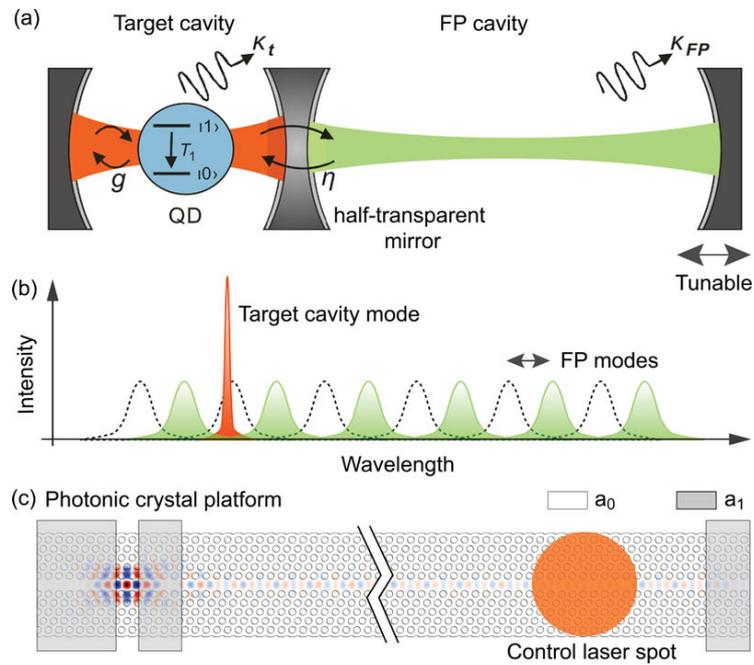

**Figure 8.** A Scheme diagram for the nonlocal control of Purcell factor. (a) Two cavities with different Q factors and mode volumes are coupled though a half-transparent mirror. (b) To bring one of the FP cavity modes into the resonance of the target cavity mode provides an extra leakage channel for the optical field confined in the target cavity. (c) A photonic crystal platform for the coupled cavities. The lattice constant along a photonic crystal waveguide is modulated in different regions to construct the target and FP cavity. A control laser beam is located in the FP cavity with a distance of ~30 μm form the target cavity.

It has been shown that the Purcell factor plays a key role in scaling down the power-density/speed limitation as well as in enhancing the optical nonlinearity in photonic switches. As seen in wavelength tuning using photonic cavities, the post-fabrication control of the Purcell factor may provide another dimension of freedom for creating novel functional nano-photonic switches. Such research has recently been triggered by the demonstration of the dynamic and static Q-factor control of photonic crystal cavities [123-125] and micro-disks [126]. Figure 8(a) depicts a coupled cavity method which is utilized for the dynamic control of the Purcell factor based on both the Q-factor and mode-volume tuning [127]. The atoms (i.e. QDs) are located at the target cavity. The Febry-Perot (FP) cavity has a lower Q-factor and larger mode volume compared to the target cavity, and a periodic mode structure is sketched in Figure 7(b). By adjusting the effective length of the FP cavity, the set of spectral modes shifts and brings one of the FP modes into resonance with the target cavity mode. This causes an extra leakage of the optical field, which has been initially confined in the target cavity, and simultaneously reduces the Q-factor and enlarges the effective mode volume of the coupled mode. Similar mechanics has been demonstrated with thermal tuning [127], photochromic tuning [128] and electromechinical tuning [129] in different kinds of coupled cavity systems. In a very recent experiment based on the above principle, a dynamic modulation of the Purcell factor and hence the spontaneous emission rate has been achieved by free carrier injections at a modulation time ~200 ps, far below the nature carrier lifetime of QDs and with a control beam ~30 μm from the target cavity [130]. This opens up the way towards the remote control of semiconductor CQED beyond GHz, leading to further novel functions in nano-photonic switches.

*4.3 Plasmonic effects*

Surface plasmons existing at the metal-dielectric interface offer an efficient means to tailor the interaction between light and matter beyond the diffraction limit [131]. Enhanced optical nonlinearity due to the presence of metallic nanostructures has been theoretically predicted, in which the localized electrical field near the metallic surface enhances the nonlinear susceptibility of the host medium [132, 133]. The theoretical framework we have discussed above indicates a local enhancement of the electrical field and the nonlinear susceptibility in proportion to the factor of $Q^2/V$. A typical plasmonic cavity can reach a Q factor of a few tens and an ultra-small mode volume of $10^{-2}$ cubic wavelength, resulting in large field enhancement that enables us to actively modulate or switch light in the vicinity of the metal nano-structures [134, 135]. To construct photonic switches, CQED systems consisting of single QDs and metallic nanowires have been proposed to utilize the "photon blockade" phenomenon in the strong coupling regime [41]. However, only static and quasi-static modulation of the properties of QDs near metallic plates has been observed so far [136, 137]. The strong optical scattering and the intrinsically high propagation loss at the metal surface, accompanied by the weak absorption of QDs, set up a barrier to demonstrate such switching devices with the current nano-technologies. Apart from conventional nano-fabrication techniques based on either chemical synthesis [138] or electron beam lithography [139], the current development on the direct growth of QDs and metal structures based on molecular beam epitaxy [140, 141] and metal-organic vapor phase epitaxy [142] provides an alternative approach in controlling the distance between QDs and metal with nanometer precision. Recently, single plasmon emission has been observed in metallic nano-structures by using chemical synthesis [143] and direct growth [144] techniques, representing a big step toward the demonstration of QD switches based on metallic nano-structures.

## 5.  Conclusion and outlook

We have discussed the fundamental limitations and the current progress of photonic switching devices based on semiconductor nanostructures. We have shown from theoretical analysis that the power-density/speed limitation of photonic switching is alleviated by the Purcell factor in photonic cavities, whilst the third-order optical nonlinearity is scaled up until the switching speed is eventually limited by the cavity photon lifetime. The current development of photonic switches shows great potential to overcome the limitations on photon number envisaged by Smith. With these prospects, nano-photonic switches are expected to meet the requirements for future optical interconnection systems between electronic chips or within chips.

However, current research on nano-photonic switches is mainly focused on single (prototype) devices and lacks a system implementation viewpoint. For instance, the reduction of operation photon number may introduce a severe degradation of the SNR [55]. There is therefore a need for further studies on the thermal performance of nano-photonic structures [145]. Moreover, the use of nano-photonic switches in high-density photonic integration needs to be carefully verified against several criteria important in practical optical logic circuits [146]. Finally the integration of nano-photonic switches with the existing Si-based electrical IC platforms is another crucial topic for future investigation [147].

**Acknowledgement**

This work is the result of contributions from many people. We are grateful to Mark Hopkinson, Osamu Kojima, Tomoya Inoue, Takashi Kita, and Kouichi Akahane for the joint research on vertical cavity QD switches, to Robert Johne, Milo Swinkels, Thang Hoang, Leonardo Midolo, Rene van Veldhoven, and Andrea Fiore for the research on ultrafast control of spontaneous emission, and to Jiayue Yuan, Matthias Skacel, Adam Urbańczyk, Tian Xia, Rene van Veldhoven, and Richard Nötzel for the collaborative work on metallic nano-structures. We would like to thank Hiroshi Ishikawa, Ryoichi Akimoto, Nobuhiko Ozaki, and Harm Dorren for stimulating discussions. We acknowledge Mark Hopkinson for a critical reading of the manuscript.